\documentstyle[12pt]{article}
\def\l{\left}
\def\r{\right}

\def\f{\frac}

\begin{document}
\baselineskip 18pt
\begin{center}
{\LARGE{Low Energy Leptogenesis in Left-Right Symmetric Models}\\}
\vskip .75cm
{\large
Avijit~Ganguly,~Jitendra~C.~Parikh~and~Utpal~Sarkar
\footnote{Address during March 1994 -- February 1995 : Institut
f\"{u}r Physik, Universit\"{a}t Dortmund,
D-44221 Dortmund, Germany.}\\
Theory Group, \\
Physical Research Laboratory, \\
Ahmedabad - 380009, India.\\
\vskip .5cm
August 1994\\}
\end{center}
\vskip .75cm
\begin{abstract}
\baselineskip 18pt
We propose a new mechanism for baryogenesis. We study the
effective potential of left-right extension of the standard
model and show that there can be a
first order phase transition at the left-right symmetry
breaking and hence $(B-L)$ symmetry breaking scale, which
is around TeV in our scenario. As a result, although $(B-L)$
violating interactions are in equilibrium at this scale, enough
$(B-L)$ asymmetry may be generated. This $(B-L)$ asymmetry is
then converted to baryon asymmetry during the anomalous
electroweak process. If right handed gauge bosons are seen in
the TeV scale, then we argue that this will be the only
consistent mechanism to generate baryon asymmetry of the universe.
\end{abstract}

\newpage
\baselineskip 18pt
In grand unified theories baryogenesis is generated through
(B+L) asymmetry of the universe
\cite{kolb, langacker}, since $(B-L)$ is either a
global symmetry (as in the case of $SU(5)$ GUT) or a local
symmetry (as in the $SO(10)$ or $E(6)$ GUT) \cite{langacker}.
The sphaleron field at high
temperature will then erase this primordial $(B+L)$
asymmetry during the electroweak phase transition.

It was then proposed that since $C$ and $CP$ are not conserved
in the standard model and there is baryon number violation due
to the anomaly\cite{hooft}, it is possible to generate enough baryon
number \cite{shapos,mstv} if the electroweak phase transition is of
weakly first order. But for this generated baryon asymmetry to
survive after the phase transition one requires that the higgs mass has
to be quite low (around 50 GeV) \cite{higgsbound}, while the
present experimental  lower bound is around 70 GeV. This makes
this interesting scenario impracticable.

Then the only viable possibility  open to us is to
generate $(B-L)$ asymmetry at higher energies, which will then
get converted to baryon asymmetry during the electroweak phase
transition in the presence of the sphaleron field, when the
$(B+L)$ violating anomalous processes are in equilibrium in
the universe \cite{fy,barlr}. Models were
proposed where the lepton number violating out-of-equilibrium
interactions of the right handed neutrinos and other scalar
particles generate the required $(B-L)$ asymmetry \cite{fy}. $C$ and $CP$
violation of correct magnitude is not a problem in such models
\cite{kiku}.

Although in the original model \cite{fy} the standard model was extended
to have a global $(B-L)$ symmetry and sterile neutrinos were added,
leptogenesis comes about more naturally in left-right symmetric
models \cite{barlr}, where $(B-L)$ is a local symmetry. Furthermore,
left-right symmetric models \cite{lr} are the most natural extension of
the standard model and are phenomenologically very attractive
particularly if they are broken at around TeV scale. In that case
in the next generation accelerators we shall see
signatures of the existence of the extra gauge
particles which mediate interactions among the right handed
particles. On the other hand, as we shall see later, the
out-of-equilibrium condition imposes a lower bound on the
left-right symmetry breaking scale \cite{mohzh}, $$ M_R \geq 10^7 \;\;\; {\rm
GeV} .$$ Will that mean that if $Z_R$ is seen in the near
future, then that will rule out the
possibility of leptogenesis completely and we have to look for
new solutions for baryogenesis ?

In this paper, we propose a new mechanism for baryogenesis through
lepton number violation in which the left-right symmetry and
hence $(B-L)$ symmetry is broken at around TeV scale \cite{fn1}.
We shall show that this phase transition can be weakly first
order. As a result, during this
phase transition bubbles of true vacuum will be formed inside
which the $CP$ non-conserving lepton number violating and hence
$(B-L)$ violating interactions will be in
equilibrium, but outside the bubbles $(B-L)$ violating
interactions can transmit from one bubble to the other along
the bubble walls only when the bubbles collide. This will not
allow the particles and antiparticles to attain equilibrium
distribution even for $M_R \sim $ TeV and hence
enough $(B-L)$ asymmetry may be
generated through the lepton  number violating interactions.
Finally during the electroweak phase
transition, this $(B-L)$ asymmetry will be converted to baryon
asymmetry of the universe via sphaleron effects.

For our purpose we consider the symmetry breaking
chain,  $
SU(3)_c \times SU(2)_L \times
SU(2)_R \times U(1)_{(B-L)}$ $ \left[ \equiv G_{LR} \right]
{M_R \atop \longrightarrow}$ $ SU(3)_c \times SU(2)_L  \times
U(1)_Y  \left[ \equiv G_{std} \right] $ $
{M_W \atop \longrightarrow}$ $ SU(3)_c \times U(1)_{em}.$
The  symmetry breaking $G_{LR}  \to
G_{std}$ takes place when the right  handed  triplet  higgs  field
$\Delta_R  \equiv$  (1,1,3,-2) acquires a vacuum  expectation
value  ({\it  vev}). In this model $(B-L)$ is a local symmetry.
The breaking of the group $G_{LR}$ also implies spontaneous
breaking of $(B-L)$. Left-right  parity implies the
existence of another
higgs field  $\Delta_L $ which  transforms  as  (1,3,1,-2)  under
$G_{LR}$.  A higgs  doublet  field  $\phi$  (1,2,2,0)  breaks the
electroweak  symmetry  and  gives  masses  to the  fermions.
With this choice of higgs scalars and usual fermions of the
standard electroweak model and the right handed neutrino (which
is always present in left-right symmetric model), the
Yukawa  couplings are  given by
\begin{equation}
{\cal  L}_{Yuk} = f_{ij}  \overline{\psi_{iL}}  \psi_{jR}  \phi
+ f_{Lij}  \overline{{\psi_{iL}}^c}  \psi_{jL}  \Delta_L^\dagger
+ f_{Rij}  \overline{{\psi_{iR}}^c}  \psi_{jR} \Delta_R^\dagger.
\label{Yuk}
\end{equation}
The scalar interactions  can be written as,
\begin{eqnarray}
V &=&
-\sum_{i,j} m^2_{ij}~tr(\phi ^{\dagger}_i
\phi_j)+\sum_{i,j,k,l}\lambda_{ijkl}[tr(\phi ^{\dagger}_i
\phi_j)~tr( \phi ^{\dagger}_k \phi_l)
+~tr(\phi^{\dagger}_i \phi_j \phi^{\dagger}_k \phi_l)] \nonumber\\
{}~&&~\nonumber\\
&&-\mu^2~(\Delta ^{\dagger}_L \Delta_L+\Delta ^{\dagger}_R
\Delta_R) +
\rho_1~[tr(\Delta ^{\dagger}_L \Delta_L)^2+tr(\Delta ^{\dagger}_R
\Delta_R)^2]\nonumber\\
&& \nonumber \\
&&+\rho_2~[tr (\Delta^{\dagger}_L \Delta_L \Delta^{\dagger}_L \Delta_L)
+tr (\Delta^{\dagger}_R \Delta_R \Delta^{\dagger}_R \Delta_R)]
+\rho_3~tr(\Delta ^{\dagger}_L \Delta_L \Delta ^{\dagger}_R
\Delta_R)\nonumber\\
{}~&&~\nonumber\\
&&+\sum_{i,j}\alpha_{ij}~(\Delta ^{\dagger}_L
\Delta_L+\Delta^{\dagger}_R \Delta_R)~{tr(\phi ^{\dagger}_i \phi_j) }
+\sum_{i,j} \beta_{ij}~[~tr(\Delta ^{\dagger}_L\Delta_L\phi_i
\phi ^{\dagger}_j) \nonumber \\
& &+tr(\Delta ^{\dagger}_R\Delta_R\phi ^{\dagger}_i
\phi_j)]+\sum_{i,j} \gamma_{ij}[~tr(
\Delta^{\dagger}_L\phi_i \Delta_R\phi^\dagger_j) +
tr(\Delta^{\dagger}_R\phi_i \Delta_L\phi^\dagger_j)]
\label{scalar}\\
\nonumber
\end{eqnarray}

\noindent where we defined, $\phi_1  \equiv  \phi~;~~~~
\phi_2  \equiv  \tau_2 \phi^{*}_1 \tau_2$.
The vacuum expectation values ($vev$) of the fields have the
following form:

\begin{eqnarray}
<\phi>&=& \pmatrix{k&0\cr 0& k^{\prime}}~~;~~
<\Delta_L>= \pmatrix{0&0\cr v_L&0}~~; \nonumber \\
<\tilde{\phi}>&=& \pmatrix{k^{\prime} &0\cr 0&k}~~;~~
<\Delta_R>= \pmatrix{0&0\cr v_R&0}~~; \nonumber
\end{eqnarray}

Following equation (\ref{Yuk}), $\Delta_{L,R}$ can decay
into two neutrinos and $\Delta^\dagger_{L,R}$ into two antineutrinos.
When the fields $\Delta_{L,R}$ acquire $vev$ and the $(B-L)$ is
spontaneously broken, there are new
$(B-L)$ number violating interactions of the fields $\Delta_{L,R}$,
$$ \Delta_{L,R} \to \phi + \phi {\hskip .5in} {\rm or} {\hskip .5in}
\phi^\dagger + \phi^\dagger $$

There are also other lepton number violating diagrams arising
from the right handed neutrino decays, when the $vev$ of
$\Delta_{R}$ gives Majorana mass to $\nu_R$. The lepton number
violating decays of $\nu_R$ are given by,
\begin{equation}
\nu_{iR}  \to l_{jL} + \bar{\phi} {\hskip .5in} {\rm or} {\hskip .5in}
 {l_{jL}}^c + {\phi} .\label{eqn1}
\end{equation}

The excess lepton number however can only be generated, if
in addition to the lepton number
violation there is  $CP$ violation. In
principle, there can be, a rephasing  invariant, complex phase in
the  Yukawa   couplings,  which  can  give  $CP$  violation.
The lepton number generation  can then occur only through
the interference of the one loop diagrams with the tree level
interactions mentioned above. The relevant tree level and one
loop diagrams are shown in fig. 1. The interference of these
diagrams can give rise to excess $(B-L)$ number if the loop
integral is imaginary .
The  magnitude  of the  asymmetry  generated  by the  left-handed
triplet $\Delta_L$ decay is given by,
\begin{equation}
\epsilon_\Delta  \approx  \frac{1}{8  \pi  |f_{Lij}|^2}
{\rm  Im}  [h^* f_{Lij}^*  f_{ik}  f_{jk}]
F \left(\frac{h^* }{f_{Rkk}}\right) , \label{eps}
\end{equation}
where, $h = \gamma_{12} + \gamma_{21}$ and
$F(q) = {\rm ln} (1 + 1/q^2)$. In general, the quantity
$[h^* f_{Lij}^*  f_{ik} f_{jk}]$ can
contain a, rephasing  invariant,  $CP$ violating phase and so can
be complex and the loop integral is imaginary.

For the right handed neutrino $\nu_R$ decays, there are two one
loop diagrams, which interfere with the tree level diagram.
The interference of the tree level diagram with the one with higgs
triplet in the loop gives a contribution equal in magnitude to
that of the triplet higgs decays. The magnitude of the lepton
asymmetry generated from the interference of the tree
level diagram with the other loop diagram is given by,
\begin{equation}
\epsilon_\nu  \approx  \frac{1}{4  \pi  |f_{ik}|^2}  {\rm  Im}  [f_{ik}
f_{il}  f_{jk}^*  f_{jl}^*]  \frac{f_{Rii}}{f_{Rkk}} .\label{eps1}
\end{equation}
The total asymmetry generated is the sum of all these
contributions and we denote it as $\epsilon$.

For the generation of the $(B-L)$ asymmetry we require another
ingredient, namely, the $(B-L)$ number generating
interactions should not  take place in equilibrium \cite{fry}. This
constraint is extremely stringent \cite{fyb}. Consider the scenario when
the right handed electron neutrino is the lightest of the
right handed neutrinos and also lighter than the triplet higgs.
In that case the $(B-L)$ asymmetry will be generated by the
$\nu_{eR}$ if it satisfies the out-of-equilibrium condition. On
the other hand if $\nu_{eR}$ does not satisfy the
out-of-equilibrium condition, then the $(B-L)$ number violating
$\nu_{eR}$ decays will erase any existing $(B-L)$ asymmetry. In
that case the sphaleron processes will erase any existing
$(B+L)$ asymmetry leaving a baryon symmetric universe causing
annihilation catastrophfe. Of course, then one has to think of
generating baryon asymmetry during the electroweak phase
transition. But with the present limit on the higgs mass such
possibilities have also been ruled out.

The out-of-equilibrium condition for the total decay rate of
the $\nu_{eR}$ can be written as,
\begin{equation}
 \mbox{} \hskip 1in \Gamma_{\nu_e}
\sim \displaystyle \frac{f_{11}^2}{16 \pi} {M_{\nu_{eR}}}
\leq 1.7 \sqrt{g} \displaystyle
 \frac{T^2}{M_{Pl}}
 \hskip .5in {\rm at} \;\;\; T=M_{\nu_e}  \label{const}
\end{equation}
For, $f_{11} \sim 10^{-5}$, we get,
\begin{equation}
M_{\nu_{eR}} \geq 10^6 {\rm GeV}  \label{const1}
\end{equation}
There are other scattering processes whose rates
should also be less than the expansion rate of the universe,
such as, $\nu_{eR} e_R \to \bar{u}_R d_R$. Combining all these
bounds, we can obtain a bound on the
$SU(2)_R$ symmetry breaking scale \cite{mohzh},
\begin{equation}
M_R \geq 10^7 {\rm GeV}.
\end{equation}
It will then mean that if we see any signature of the right
handed gauge bosons at energies much below this scale, then we
have to think of new mechanisms for the generation of the baryon
asymmetry.

In this article we argue that for the choice of higgs scalars
we have in this model, it is possible to make the left-right
symmetry breaking phase transition to be first order and hence
generate $(B-L)$ asymmetry during this symmetry breaking even
if the symmetry breaking scale is around TeV and does not
satisfy the out-of-equilibrium condition. For this purpose we
show that after intregating out the quadratic part, the finite
temperature effective potential for this
model at the one loop level contains linear terms in $T$, which
after minimization can make this phase transition a first order
\cite{choi}.

For this purpose we will  introduce thermal field theory and the
concept of thermal effective action in the imaginary time formalism
of Matsubara.
We compute the partition function $Z$ and change the time
coordinate $t$ over to a variable defined as $ \tau = i t $ with the limits
of integration varying between  0 to $ \beta $ and perform the functional
integration over the fields (e.g $\phi(\tau, x)$) with a periodic
(antiperiodic) boundary conditions in $ \tau $ for bosons (fermions).


For a Lagrangian that is quadratic in the field variables, one can
compute the partition function Z exactly by expanding the field
variables e.g $ \phi (t,x) $ as
\begin{equation}
\phi \l (\beta, x \r ) = {\frac {1}{\beta} } {\sum_n} \int
{\frac {d^3 p}{(2\pi)^3}} e^{-i(\omega_n \tau - p .x )}
\tilde{\phi \l(\omega_n ,p \r)}
\end{equation}
and performing the Gaussian integration over the field
variables.  Here $ \omega_n = {\frac {2 \pi n}{\beta} } (n=
- \infty~~{\rm{to}}~~ \infty )$
are the (Matsubara) frequencies for bosons and  have  been defined to agree
with the periodic boundary conditions of the field variables but
otherwise one has to perform expansion and compute the
quantities to a certain order in perturbation theory.
Starting from the partition function $Z$, we define the
effective action in the presence of a constant background field
$\overline{\Gamma} \l (\phi_c \r )$  in terms of the Legendre
transform \cite{bern} of the generating functional for the connected Green's
function $ W^{\beta} \l(J \r )  = \ln Z^{\beta} \l(J \r ) $ as
\begin{equation}
\overline{\Gamma} \l (\phi_c \r ) = W^{\beta} \l[ J \r] -
\int d {\bar{x}} \phi_c \l( \bar{x} \r) J \l( x \r)
\end{equation}
where $\phi_c \l(\bar x \r)$ is the classical field defined as
$\phi_c \l(\bar{x} \r) = \frac {\delta W^{\beta} \l [J \r ] }
{\delta J \l [ \bar{x} \r ] } $ and the source is given by
$ J \l(\bar{x} \r) = -\frac {\delta {\Gamma \l [\phi_c \r]}}{\delta
\phi_c \l[\bar{x} \r]}
$, where the vector $\bar{x} = \l(- i \tau ,\vec{x} \r) $.

The quantity $\overline{\Gamma} \l (\phi_c \r ) $, evaluated
semiclassically  about some field configuration $\phi_c
\l( \bar{x} \r) $,  gives the
free energy of the system in that configuration.
The effective potential $V_{eff} \l( \phi \r)$, which is
the first term in a derivative expansion of $\overline{\Gamma}
\l (\phi \r) $, is  the free energy density in a
background constant field configuration. We shall study
the effective potential for determining the order
of the phase transition for the model under consideration.  In
the case of a first order phase transition, usually
signalled by the presence of a  $\phi^{3}$ term
in the effective potential, a system can be
trapped temporarily in a metastable state leading to
non-equilibrium phenomena.  The rate of decay for such a system
is determined from the imaginary part of its free energy
\cite{aff} or effective potential.
We have investigated this in the one loop approximation at high temperature.
With the choice of Higgs fields
under consideration, the  Lagrangian is given by $$ { \cal L} =
{\cal L}_{gauge field} + {\cal L}_{Higgs} + {\cal L}_{fermions} .
$$  The gauge field part of the lagrangian contains the kinetic
energy terms for the gauge bosons corresponding to the gauge groups
$SU(2)_L \times SU(2)_R \times U(1)_{B-L}$. The gauge coupling
constants for the gauge groups $SU(2)_L$ and $SU(2)_R$ are same
and we denote it by $g$, while that of the $U(1)_{B-L}$
is denoted by $g'$. The higgs part of the
lagrangian contains the kinetic energy terms for the fields $\Delta$s
and the field $\phi$, and the scalar interaction terms given by
equation \ref{scalar}. The fermionic part of the lagrangian contains
the kinetic energy terms for the fermions and the Yukawa
couplings given by equation \ref{Yuk}.
The symmetry breaking scales that we consider are
$ \langle \phi \rangle :
(k^2+ {k'}^2)^{\frac {1}{2}} \simeq 250 GeV $;~~~$
\langle \Delta_R \rangle : { v_R} \simeq
1 - 10  TeV $ and $\langle \Delta_L \rangle : { v_L} \simeq
10^{-1} - 10^{-5}  GeV $.

We have computed the effective potential at the one loop level,
in the Feynman gauge
by integrating out the quadratic field variables present
in the Lagrangian. Upon doing so we have arrived at  integrals
of the form
$$
I_{\pm} \l( k \r) = \int_{0}^{\infty} dk {k^2} \ln \l[1 \mp e^{\beta\sqrt{
k^2+M^2}} \r]
$$
where $I_{+}$ is used for bosons and $I_{-}$ is used for fermions.
 From here, following Dolan and Jackiw \cite{doljk},
we have evaluated these integrals in the high temperature limit.
In this limit the effective potential can be written as
 $$ V = V_{tree} + V_{T=0} + V_T $$
Where $$
  V_T \simeq A {\frac  {{\pi}^2 T^4}{90} + B {v_R}^2 T^2 -C T
+D {\rm ln} {(v_R T)}^2 + E}
$$
and A, B, C, D and E are the functions of the parameters present
in the theory. The first term is from the familiar ultra-relativistic
limit of a gas of bosons or fermions, the other terms are the corrections
to it, coming from the interactions present in the theory. The
term E is the one containing  zero temperature effects.
The term linear in T is the one we are interested
in, because that is the one responsible for first order phase
transition, and it is given by
$$ C  = {\frac
{2\pi}{3}} \left[ \left( m_{Z_L}^3+m_{Z_R}^3+2{m_{W_L}^3}+2{m_{W_R}^3}
\right)
+\f{1}{16\pi} \left( 3M^3+2{M_{R1}}^3+{M_{R2}}^3 \right) \right.
$$  $$ \left.
+\f{1}{8\pi} \left( {\Sigma_{a=1}}^3 {M_{\phi a}}^3 \right) \right]$$

Here $$
{M_{R1}}^2= {\mu_{\Delta_L}}^2+2{\bar{\lambda_R}} {v_R}^2
$$
$$
{M_{R2}}^2= {\mu_{\Delta_L}}^2+4{\bar{\lambda_R}} {v_R}^2
$$
$$
{M}^2= {\mu_{\Delta_L}}^2+{\bar{\lambda_R}} {v_R}^2
$$
and
$$
{M_{\phi_1}}^2= \left[ m^2+3{\bar{\lambda_\phi}} \left( k^2+k^{'2}\right)+
2{\bar{\lambda_{LR}}{v_R}^2}+2 \left(
\bar{\lambda_{LR}^1}{v_R}{v_L} \right) \right]
$$
$$
{M_{\phi_2}}^2= \left[ m^2+3{\bar{\lambda_\phi}} \left( 4k^2+2k^{'2}\right)+
2{\bar{\lambda_{LR}}{v_R}^2}+2 \left(
\bar{\lambda_{LR}^1}{v_R}{v_L} \right) \right]
$$
$$
{M_{\phi_3}}^2= \left[ m^2+3{\bar{\lambda_\phi}} \left(2k^2+4k^{'2}\right)+
2{\bar{\lambda_{LR}}{v_R}^2}+2 \left(
\bar{\lambda_{LR}^1}{v_R}{v_L} \right) \right]
$$
and $$ \bar{\lambda_\phi}= -\lambda_{\phi}$$

where, we have defined $\lambda_L = (\rho_1 + \rho_2)$;
{\underline{$\lambda_\gamma = \gamma_{12} + \gamma_{21}$}} and $\lambda_{LR}
= (\alpha_{12} + \alpha_{21} + \beta_{12} + \beta_{21})$.
 For a range of choice of parameters ( $\lambda$'s are usually
taken to be
less or equal to one)
the Left-Right symmetry breaking phase transition can be made
to be of first order. It is worth pointing out here that since
the symmetry breaking scale for the right handed particles is
greater than all the other mass scales present in the theory, this one
loop result has an imaginary piece in it, that signifies decay of
the system from symmetric  to the asymmertic vacuum through tunneling.
This decay rate is proportional to the imaginary part of the
effective potential. One can improve the result of this analysis
in the infrared region by summing up the daisy diagrams but
according to the standard folklore that only introduces a factor
of $\f {2}{3}$ in front of the cubic term, and hence makes the transition
weakly first order. Some of these studies (both numerical and analytical)
are currently under consideration.

In the $(B-L)$ broken phase, most of the fermionic fields are
massless. The source of $CP$ violation in this phase is through the Majorana
mass terms of the right handed neutrinos \cite{fy,barlr,kiku}.
As it has been mentioned
previously, in this sector there is enough $CP$ violation and
the amount of lepton number asymmetry generated is given
by $$\langle {\cal O} \rangle = \displaystyle \frac{2 \epsilon_\Delta
+ \epsilon_\nu}{g_*},$$  where, $g_*$ is the effective numbers
of massless degrees of freedom and after the left-right symmetry
breaking it is $g_* \sim 400$; To obtain the amount of
lepton number asymmetry inside this domain wall one has to solve the
Boltzmann equation. However, in the limit of large time, the
asymptotic order of magnitude for the solution of
the Boltzmann equation can be estimated following earlier references
\cite{fy} to be as given by the above expression. On the other hand,
in the $(B-L)$
conserving phase there is no CP violation since all fermion masses
(including the Majorana masses of the right handed neutrinos)
vanish. In this phase, one can then rephase the
fermion fields to absorb all the phases. We denote the thickness
of the domain walls which separates these two phases by $l$,
which is related to the inverse of the time derivative of the
scalar operator for the field $\Delta_R$. The detail
calculation of the sphaleron transition probability between
the two phases will depend on the wall thickness.
An order of magnitude estimate for the lepton
number asymmetry thus generated is given by \cite{mstv},
$$ \displaystyle \frac{n_{L}}{s} \approx \frac{1}{g_*}
\left( \frac{g^2}{4 \pi} \right)^4 l \langle {\cal O} \rangle. $$
The observed baryon asymmetry is then generated
from this $(B-L)$ asymmetry during the electroweak phase
transition, during the epoch when the $(B+L)$ number violating
interactions are in equilibrium in the universe, and is given by
\cite{ht},
$$ n_B = - \left( \displaystyle \frac{8
N_g + 4 N_H}{22 N_g + 13 N_H} \right) n_{(B-L)} \approx
\frac{1}{3} n_L. $$ For a
wide range of choice of parameters this can give the required
amount of baryon asymmetry, since the parameters in the left-right
symmetric model are not constrained by phenomenology. Consider
the expression for the $\epsilon$s, where only the Yukawa couplings
of the right handed neutrinos enter. But both the Majarana as well
as the Dirac masses of these particles are free parameters and one
can choose a large combination of these parameters, which can
reproduce the amount of baryon asymmetry required by the standard
big-bang model.

To summarize, we have studied the effective potential of the left-right
symmetric model at finite temperature and have shown that the left-right
symmetry breaking can take place through first order phase transition
at around TeV scale. This can then generate enough $(B-L)$ asymmetry,
which can then generate enough baryon asymmetry during the
electroweak phase transition.

{\bf Acknowledgement} One of us (US) would like to acknowledge a
fellowship from the Alexander von Humboldt Foundation and hospitality
from the Institut f\"{u}r Physik, Univ Dortmund during his research
stay, where part of this work was done.

\vskip .5in
\noindent
\subsection*{Figure Caption}

{\bf Figure 1} Diagrams contributing to the generation of lepton asymmetry.

\end{document}